# Coupling Lattice Distortion and Cation Disorder to Control Li-ion Transport in Cation-Disordered Rocksalt Oxides

Zichang Zhang[1,#], Lihua Feng[2,#], Jiewei Cheng[1], Peng-Hu Du[1], Chu-Liang Fu[2], Jian Peng[2,*], Shuo Wang[2,*], Dingguo Xia[1,3], Xueliang Sun[2], and Qiang Sun[1,3,*]

[1]*School of Materials Science and Engineering, Peking University, Beijing, 100871, China*

[2]*Zhejiang Key Laboratory of All-Solid-State Battery, Eastern Institute of Technology, Ningbo, 315200, China.*

[3]*Beijing Key Laboratory of Theory and Technology for Advanced Batteries Materials, Peking University, Beijing 100871, China*

*Corresponding author   *Email address:* sunqiang@pku.edu.cn (Q. Sun); shuowang@eitech.edu.cn (S. Wang); jpeng@eitech.edu.cn (J. Peng)

[#]The authors (Z. C. Zhang and L. H. Feng) contribute equally to this work.

## Abstract

Cation-disordered solids offer a rich chemical landscape in which local coordination, lattice responses, and configurational disorder collectively—but often implicitly—govern ion transport. In cation-disordered rocksalt oxides, Li$^+$ diffusion has conventionally been rationalized using the static 0-transition-metal (0-TM) percolation rule, which treats the lattice as an ideal, passive framework and is therefore intrinsically limited in capturing experimentally accessible capacities. Here, we show that lattice distortion constitutes an essential and previously overlooked independent chemical degree of freedom that actively reshapes ion-percolation networks in disordered oxides. By developing a lattice-responsive computational framework that integrates Monte Carlo sampling of cation configurations with molecular dynamics simulations accelerated by machine-learning interatomic potentials, we quantitatively resolve Li$^+$ percolation networks and electrochemical capacities, with deviations from experimental values below 5%. Our results reveal a causal coupling between lattice distortion and cation short-range order: enhanced local distortions precede and suppress short-range ordering and activate Li$^+$ migration through nominally inaccessible 1-transition-metal (1-TM) diffusion channels, fundamentally extending percolation beyond the conventional 0-TM paradigm. Guided by this chemical mechanism, we design and synthesize a high-entropy cation-disordered oxide, $Li_{1.2}Mn_{0.2}Ti_{0.2}V_{0.2}Mo_{0.2}O_2$, which exhibits enhanced lattice distortion and achieves an improved Li$^+$ percolation network of 71.9%, surpassing the 65.8% observed in the reference material $Li_{1.2}Mn_{0.4}Ti_{0.4}O_2$. This enhanced percolation enables a high experimental capacity of 256.3 mAh/g, in close agreement with our predicted value of 255.1 mAh/g. These findings establish lattice distortion as an active chemical control parameter for ion transport in disordered solids, revising prevailing percolation concepts and providing a general, chemically grounded design principle applicable beyond metal-ion cathodes.

## 1. Introduction

Cation-disordered solids occupy a unique position in solid-state chemistry, combining chemical simplicity with profound structural and functional complexity[1]. Among them, cation-disordered rocksalt (DRX) oxides have attracted sustained attention as lithium-ion battery cathodes owing to their ability to deliver high capacities using earth-abundant transition metals, circumventing the reliance on scarce and costly elements such as Ni and Co[2,3]. Beyond their technological relevance, DRX oxides provide a chemically rich platform for exploring how local atomic environments, disorder, and lattice responses collectively govern ion transport in solids[4,5]. In contrast to conventional layered cathodes, where Li$^+$ and transition-metal (TM) ions occupy well-defined crystallographic sublattices, DRX materials exhibit a stochastic distribution of Li$^+$ and TM cations over octahedral sites within a close-packed oxygen framework (Fig. 1a)[6]. This disordered cationic arrangement gives rise to several distinctive features, including high accessible capacities enabled by excess lithium and anionic redox activity, isotropic and mechanically robust lattice responses during cycling, and broad compositional flexibility that accommodates a wide range of earth-abundant elements[7–9]. These characteristics make DRX oxides not only promising cathode materials but also prototypical systems for studying ion transport in chemically disordered lattices.

A central challenge in understanding and designing DRX cathodes lies in rationalising Li$^+$ transport through the disordered lattice. Lithium migration proceeds via octahedral–tetrahedral–octahedral pathways, with the local configuration of neighbouring TM cations governing the migration barrier through steric and electrostatic interactions. The long-range connectivity of low-barrier diffusion motifs defines the percolating Li$^+$ network and ultimately determines the fraction of lithium that can be reversibly accessed[10]. Establishing reliable descriptors of Li$^+$ percolation is therefore essential for connecting local chemical environments to macroscopic electrochemical behaviour.

Foundational work by Ceder and co-workers introduced the 0-TM percolation rule, which classifies diffusion channels based on the number of TM cations surrounding the intermediate tetrahedral site[2,3,6]. Within this framework, only tetrahedra free of neighbouring TM ions (0-TM channels) are considered active, while 1-TM and higher-order environments are assumed to be blocked by prohibitive electrostatic repulsion[11,12]. This picture has provided an influential and conceptually clear basis for analysing short-range order and Li$^+$ connectivity in DRX materials. However, percolation models built on the static 0-TM criterion consistently underestimate experimentally accessible capacities[12]. For example, the predicted Li$^+$ percolation fraction is zero for Li$_{1.2}$Mn$_{0.4}$Zr$_{0.4}$O$_2$ (LMZO) and only ~35% for Li$_{1.2}$Mn$_{0.4}$Ti$_{0.4}$O$_2$ (LMTO), in stark contrast to experimentally observed values of 43.3% and 65.8%, respectively[12]. This systematic discrepancy highlights a fundamental limitation of percolation descriptions that rely solely on idealised geometric topology.

Ion transport in disordered oxides is inherently sensitive to the local energy landscape, which is shaped not only by cation configuration but also by lattice distortions and atomic displacements[13,14]. Experimental and theoretical studies have shown that deviations from ideal rocksalt geometry, such as oxygen sublattice distortions and off-centering of TM ions, can substantially modify Li$^+$ migration barriers[2]. Incorporating experimentally derived distortions has been shown to partially activate nominally blocked diffusion channels and extend Li$^+$ connectivity in specific DRX compositions[15]. These observations indicate that lattice distortion plays a critical role in Li$^+$ kinetics. Yet, the extent to which distortion quantitatively reshapes the global percolation network, and how it couples to cation short-range order, remains unresolved. More broadly, existing models lack a framework that treats the lattice as a responsive chemical entity rather than a static scaffold.

Here we demonstrate that lattice distortion constitutes an essential and previously unaccounted-for degree of freedom in Li$^+$ percolation within cation-disordered oxides. We develop a lattice-responsive computational framework that integrates Monte Carlo sampling of cation configurations with molecular dynamics simulations accelerated by machine-learning interatomic potentials, enabling simultaneous treatment of configurational disorder and structural relaxation. This approach quantitatively reproduces experimentally measured percolation fractions and capacities with deviations below 5%, resolving long-standing discrepancies between theory and experiment.

Beyond quantitative agreement, our results reveal a causal coupling between lattice distortion, cation short-range order, and Li$^+$ transport. Enhanced local distortions suppress short-range ordering and activate diffusion through nominally inaccessible 1-TM channels, fundamentally expanding the percolation landscape beyond the conventional 0-TM paradigm. Guided by this mechanistic insight, we design and synthesise a high-entropy DRX oxide, $Li_{1.2}Mn_{0.2}Ti_{0.2}V_{0.2}Mo_{0.2}O_2$, which exhibits increased lattice distortion, an expanded percolating Li$^+$ network, and a high reversible capacity in close agreement with theoretical predictions. By establishing lattice distortion as an active chemical control parameter for ion percolation, this work revises the prevailing understanding of transport in disordered oxides and provides a general design principle applicable to a broad range of ion-conducting materials beyond lithium-ion cathodes.

## 2. Results and Discussion

### 2.1 The effects of local lattice distortion in Li$^+$ diffusion.

In the DRX structure, the random occupation of octahedral sites by Li$^+$ and TM ions creates distinct local environments, as determined by the arrangement of Li and its neighboring TM ions (Fig. 1a). When centered on the intermediate tetrahedral site in the O-T-O diffusion pathway, the surrounding cation tetrahedral clusters can be classified into 0-, 1-, 2-, 3-, and 4-TM configurations. Generally, only 0-TM diffusion channels are considered viable for Li$^+$ conduction, as neighboring TMs in 1-TM and 2-TM configurations are thought to induce strong Coulombic repulsion (Fig. 1a), resulting in a high energy barrier. In contrast, 3-TM and 4-TM configurations lack available Li sites for hopping and are therefore incapable of supporting Li$^+$ diffusion.

Strategic selection of cationic elements enables tuning of the Li$^+$ percolation network by optimizing cation configuration. One representative example is LMTO, which exhibits a significantly enhanced percolation network compared to LMZO, despite their identical composition and average TM valence. We employ special quasi-random structures (SQSs) to model these structures, which effectively capture the random cation distribution characteristic of DRX cathodes[16]. Furthermore, we evaluate the Li$^+$ diffusion barriers in 0-TM and 1-TM diffusion channels, as shown in Fig. 1b. The barriers follow the trend: 0-TM (0.30 eV) < 1-TM (Mn) (0.41 eV) < 1-TM (Ti) (0.48 eV) ≈ 1-TM (Zr) (0.49 eV), reflecting increasing Coulombic repulsion between the diffusing Li$^+$ ion and neighboring cations (Li$^+$, Mn$^{3+}$, Ti$^{4+}$, or Zr$^{4+}$). Notably, some low-barrier 1-TM channels exhibit barriers comparable to those of 0-TM channels, whereas 2-TM diffusion channels display substantially higher barriers (>1.0 eV), as detailed in Supplementary Note 1.

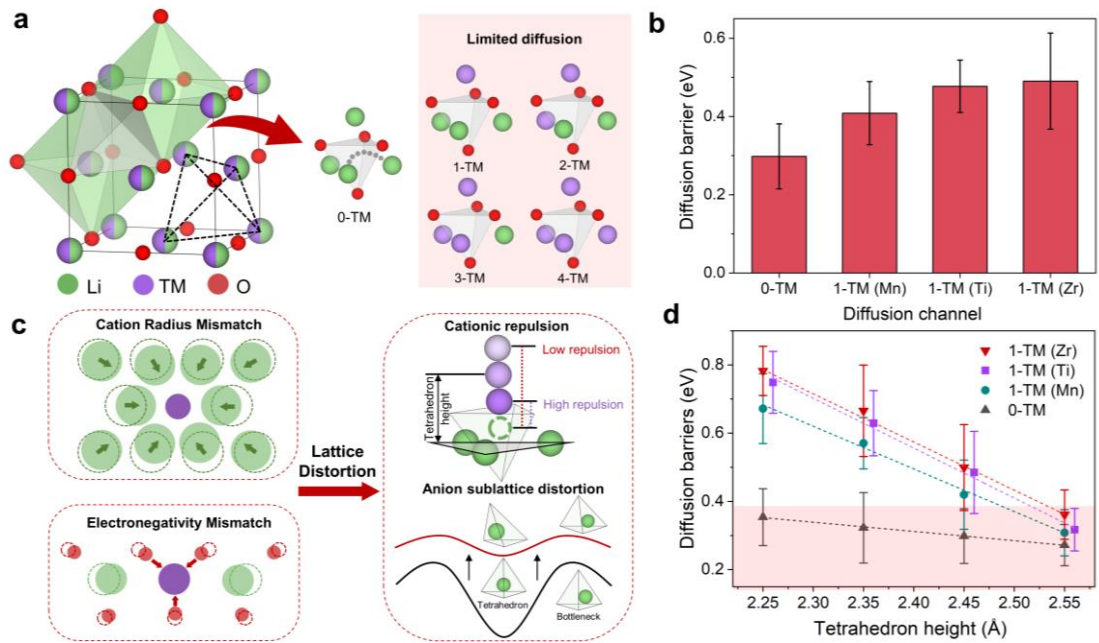

**Fig. 1. The effects of lattice distortion in Li$^+$ diffusion.** (a) The DRX cathode structure and five types of tetrahedral clusters within the structure. Pink shading marks the n-TM (n ≥ 1) clusters, which contain one or more TM cations and impede Li$^+$ diffusion. (b) The calculated energy barriers of Li$^+$ diffusion through 0-TM and 1-TM tetrahedral channels. (c) The Li$^+$ diffusion energy landscape modulated by lattice distortion (cationic repulsion and anion sublattice distortion). (d) The calculated Li$^+$ diffusion barriers through 0-TM and 1-TM pathways with various tetrahedron heights.

The conventional 0-TM percolation rule is overly simplistic, as it neglects lattice distortion. This distortion originates from chemical unevenness within the crystal lattice, primarily driven by mismatches in cation radii and electronegativities[17,18]. Such heterogeneity induces oxygen sublattice distortion through uneven TM–O and Li–O bonding, causing cations to deviate from their ideal octahedral centers. As a result, the Li$^+$ migration energy landscape is profoundly reshaped through Coulombic repulsion with off-center TM cations and bonding interactions with oxygen anions (Fig. 1c)[15,19]. Quantifying the effects of lattice distortion is expected to narrow the gap between predicted and experimental Li$^+$ percolation. The tetrahedral cluster height, defined as the vertical distance between the triangular base and the fourth atom (Fig. 1c), is employed as a descriptor of distortion extent. The Li$^+$ migration barriers of 0-TM and 1-TM in various tetrahedral heights (2.25 Å, 2.35 Å, 2.45 Å, and 2.55 Å) are calculated

(Fig. 1d). Notably, as 0-TM diffusion barriers remain low and relatively insensitive to height variations, the 1-TM diffusion barriers show significant dependence on the heights. For example, in the 1-TM (Mn) channel, the barrier reduces dramatically from 0.67 eV to 0.31 eV as the tetrahedral height increases from 2.25 to 2.55 Å, rendering the diffusion pathway activated and capable of delivering additional capacity. The critical tetrahedral heights ($H_{tetra}$) required to achieve a diffusion barrier below 0.4 eV (a typical value within the 0-TM barrier range[15]) are determined to be 2.476 Å (Mn), 2.501 Å (Ti), and 2.522 Å (Zr) in their respective 1-TM clusters.

## 2.2 Enhancing the percolation network through local lattice distortion.

The macroscopic electrochemical capacity of DRX cathodes is governed by the connectivity of active Li sites through low-barrier diffusion pathways, which constitute the Li percolation network[12]. However, the 0-TM percolation rule systematically underestimates experimental capacities. For example, LMZO delivers a first-cycle capacity of 141.3 mAh/g, whereas the 0-TM percolation rule predicts no percolation[12]. Similarly, LMTO exhibits an experimental capacity of 260.0 mAh/g, significantly exceeding the 0-TM percolation rule prediction of 138.3 mAh, with an underestimation of 46.8 %. This discrepancy arises from neglecting diffusion channels activated by lattice distortion (Fig. 1c). While the 0-TM percolation rule yields isolated Li clusters incapable of percolation, local lattice distortions can activate 1-TM channels, effectively connect these clusters, and dramatically enhance Li$^+$ percolation (Fig. 2a).

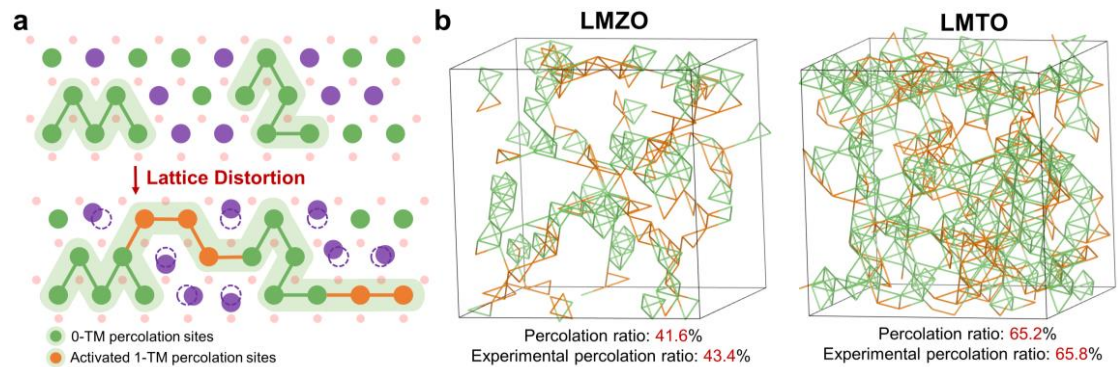

**Fig. 2. The effects of lattice distortion in Li$^+$ percolation.** Green and orange represent 0-TM and 1-TM diffusion channels, respectively. (a) The Li$^+$ percolation network activated by local lattice distortion. For clarity, the displacements of O and Li atoms are not illustrated. (b) Percolation

network for LMZO and LMTO obtained from hybrid MC-MD simulations.

The Li$^+$ percolation network can be rationally engineered by tailoring cation composition[20]. For example, the cation radius mismatch between the TM$^{4+}$ and Li$^+$ is significantly larger in LMTO (Ti$^{4+}$: 0.605 Å; Li$^+$: 0.76 Å) than in LMZO (Zr$^{4+}$: 0.72 Å; Li$^+$: 0.76 Å). The enhanced size mismatch is expected to amplify the lattice distortion, thereby reshaping the intermediate tetrahedral sites and flattening the Li-ion diffusion energy landscape (Fig. 1c). Consequently, LMTO is expected to enable more extensive percolation compared to LMZO, a trend consistent with experimental observations of an enhanced percolation network in LMTO[12].

To test this hypothesis, we developed a hybrid MC-MD model to track the evolution of Li$^+$ percolation networks in LMZO and LMTO. The MC component samples thermodynamically stable cation configurations, while the MD component captures local lattice distortions, enabling the simultaneous sampling of both configurational and lattice degrees of freedom. The MC-MD simulations are accelerated by MLFFs built by fine-tuning CHGNet[21] (details in Supplementary Note 2). By incorporating 1-TM diffusion channels with $H_{\text{tetra}}$ exceeding the critical thresholds (Mn: 2.476 Å; Ti: 2.501 Å; Zr: 2.522 Å) into the percolation network, both LMZO and LMTO exhibit significantly enhanced percolation compared to the results of 0-TM diffusion channels alone (Fig. 2b). Specifically, the percolation ratio increases from 0 to 41.6% in LMZO and from 43.3% to 65.2% in LMTO (Fig. 3b), with the corresponding predicted and experimental capacities summarized in Table 1. Remarkably, the predicted capacities and percolation ratios show excellent agreement with experimental values, with an error of less than 5%[12]. As expected, substituting Zr$^{4+}$ with a smaller Ti$^{4+}$ cation enhances the local lattice distortion, increasing the TM ion displacement from 0.16 Å in LMZO to 0.20 Å in LMTO (Fig. 3c and Fig. S6), and enables an enriched three-dimensional percolation network by activating more 1-TM pathways.

**Table 1.** Fraction of Li based on 0-TM and modified 0-TM percolation rule after hybrid MC-MD simulations, with comparison of the first-cycle capacity predicted by simulation and experiments.

| Cathodes | Fraction of Li (0-TM) | Fraction of Li (0-TM+1-TM) | Fraction of Li (Exp.) | Capacity (Pred., mAh/g) | Capacity (Exp., mAh/g) | Ref |
|---|---|---|---|---|---|---|
| LMZO | 0% | 41.6% | 43.4% | 135.5 | 141.3 | 12 |
| LMTO | 43.3% | 65.2% | 65.8% | 257.5 | 260.0 | 12 |
| LMTVMO | 60.2% | 71.6% | 71.9% | 255.1 | 256.3 | This work |

## 2.3 Screening High-Entropy DRX cathodes with enhanced percolation.

The high-entropy strategy has been widely employed to diversify the local environment and engineer a frustrated energy landscape that promotes ionic conduction[14,22]. Here, we leverage the high-entropy strategy to DRX to tune the Li$^+$ percolation network by engineering lattice distortion. To ensure synthetic feasibility, we select LMTO as the host and explore the composition Li$_{1.2}$Mn$_{0.2}$Ti$_{0.2}$M$^{(I)}_{0.2}$M$^{(II)}_{0.2}$, where M$^{(I)}$ and M$^{(II)}$ denote 17 distinct TMs (Fig. 3a), yielding a total of 289 SQSs.

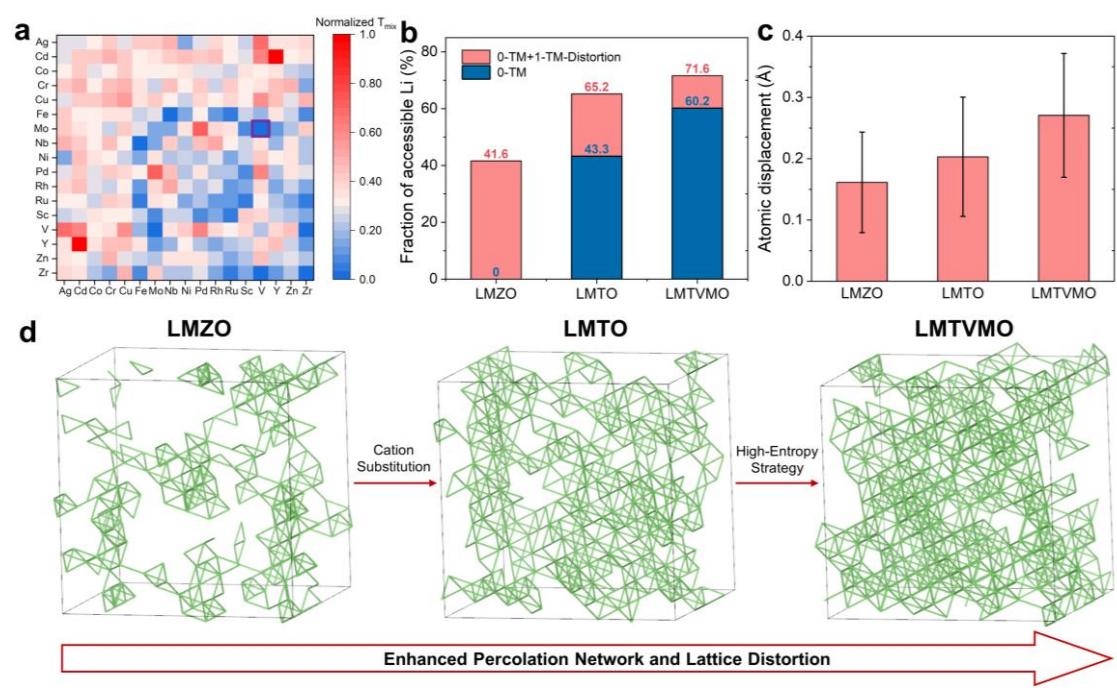

**Fig. 3. Screening of high-entropy DRX cathodes and percolation analysis.** (a) Mixing temperatures of Li$_{1.2}$Mn$_{0.2}$Ti$_{0.2}$M$^{(I)}_{0.2}$M$^{(II)}_{0.2}$O$_2$ in different M$^{(I)}$ and M$^{(II)}$ combinations. The color bar indicates an increase in incompatibility from blue to red. The purple square labels the most promising high-entropy DRX cathode. (b) Fraction of accessible Li in different cathodes, calculated using only the 0-TM percolation rule or including activated 1-TM diffusion channels. Each case takes an average of over 500 configurations obtained from hybrid MC-MD simulations. (c) The average atomic displacement of TMs in MC-MD converged LMZO, LMTO, and LMTVMO, with standard deviation represented by an error bar. (d) Percolation network of LMZO, LMTO, and LMTVMO obtained from hybrid MC-MD simulations.

We calculated the mixing temperature of candidates (Fig. 3a), which represents the thermal energy required to mix the TMs, as a qualitative measure of the synthetic accessibility of $Li_{1.2}Mn_{0.2}Ti_{0.2}M^{(I)}_{0.2}M^{(II)}_{0.2}O_2$. The $Li_{1.2}Mn_{0.2}Ti_{0.2}V_{0.2}Mo_{0.2}O_2$ (LMTVMO) exhibits the lowest mixing temperature of 1712 K, which is lower than that of LMZO (2029 K) and LMTO (2226 K), indicating its synthetic feasibility[23]. This trend aligns with previous reports that high-entropy DRX cathodes generally exhibit lower mixing temperatures than their low-entropy counterparts[24], suggesting LMTVMO has greater configurational disorder than LMTO at equivalent temperatures.

The average atomic displacements from the ideal lattice are calculated to characterize the lattice distortion. We found that the incorporation of V and Mo in LMTVMO would intensify the lattice distortion with enhanced TM deviation displacement from 0.20 Å in LMTO to 0.27 Å (Fig. 3c). This enhanced distortion reduces the coulombic repulsion between TM ions and nearby Li sites and activates additional 1-TM $Li^+$ diffusion channels (Fig. S6). Consequently, a percolation fraction of 71.6% is achieved in LMTVMO (Fig. 3b), marking a clear improvement from 65.2% in LMTO. These results underline that tuning chemical unevenness through cationic replacement and the high-entropy strategy is an effective route to expand the $Li^+$ percolation network in DRX cathodes (Fig. 3d and Supplementary Note 3).

## 2.4 Experimental verification of $Li_{1.2}Mn_{0.2}Ti_{0.2}V_{0.2}Mo_{0.2}O_2$.

Guided by the theoretical screening and percolation analysis, a prototype DRX cathode material, $Li_{1.2}Mn_{0.2}Ti_{0.2}V_{0.2}Mo_{0.2}O_2$ (LMTVMO), was successfully synthesized via a one-pot mechanochemical method, enabling rapid and scalable preparation of the high-entropy composition. Scanning electron microscopy (SEM) and high-resolution transmission electron microscopy (HRTEM) analyses reveal that the as-synthesized LMTVMO particles possess sizes in the range of approximately 200–

500 nm (Fig. 4a and Fig. S8). The corresponding selected-area electron diffraction (SAED) patterns (Fig. 4b) exhibit well-defined diffraction rings, confirming the polycrystalline nature of the particles observed in Fig. 4a. Synchrotron X-ray diffraction (XRD) measurements further confirm the formation of a single-phase cation-disordered rocksalt structure without detectable impurity phases (Fig. 4c). Rietveld refinement indicates that the LMTVMO crystallizes in the cubic Fm-3m space group, with a refined lattice parameter of $a$ = 4.16040 Å (Supplementary Table S2). The homogeneous distribution of constituent elements was examined by transmission electron microscopy coupled with energy-dispersive X-ray spectroscopy (TEM/EDS). As shown in Fig. 4d, representative elemental mapping images demonstrate a uniform spatial distribution of all TMs as well as oxygen throughout the particles, evidencing successful atomic-scale mixing consistent with a high-entropy DRX framework.

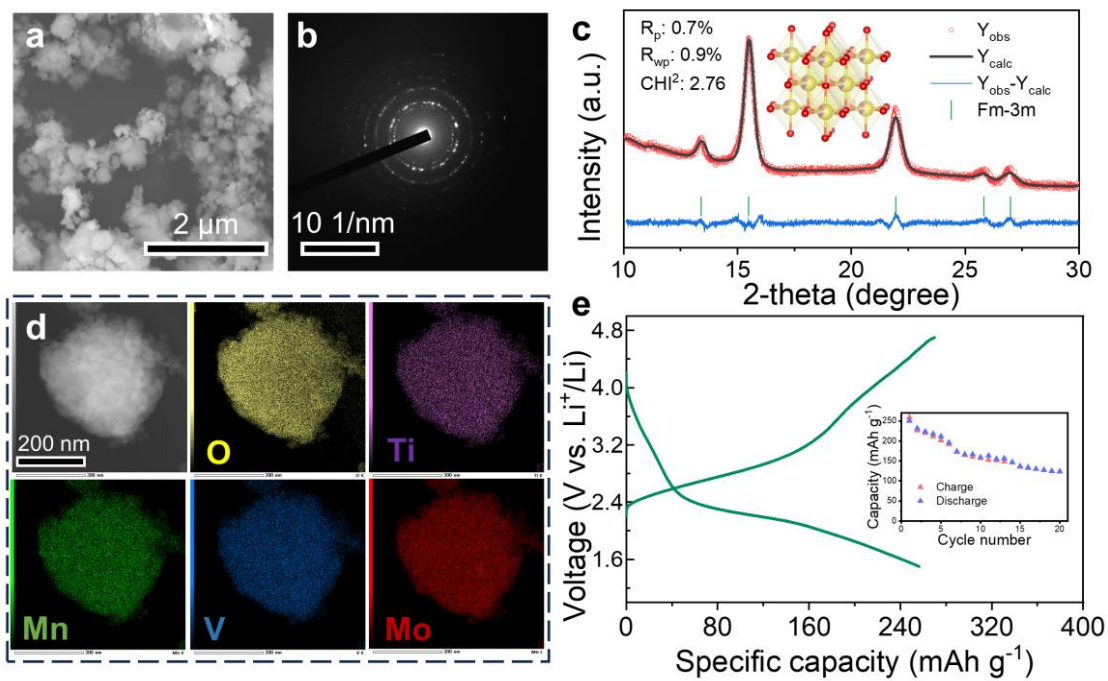

**Fig. 4. Structural characterization and electrochemical performance of the as-synthesized LMTVMO materials.** (a) SEM images of LMTVMO. (b) TEM electron diffraction patterns of LMTVMO. (c) Synchrotron XRD patterns and refined lattice constants of LMTVMO at wavelength λ= 0.56025 Å. (d) STEM/EDS mapping of elemental distribution in a particle cluster of LMTVMO. (e) Voltage profiles of LMTVMO within the voltage window of 1.5–4.7 V at 10 mA/g and room temperature. The corresponding capacity retention plots are shown as insets.

The electrochemical performance of LMTVMO was evaluated using galvanostatic cycling and compared with reference compositions. As shown in Fig. 4e, when cycled within a voltage window of 1.5–4.7 V at a current density of 10 mA/g, LMTVMO delivers a high initial discharge capacity of 256.3 mAh/g. Notably, this value closely matches the theoretically predicted capacity of 255.1 mAh/g derived from the hybrid MC–MD percolation framework, providing strong experimental validation of the lattice-distortion-activated percolation mechanism proposed in this work. As summarized in Table 2, the electrochemical performance of LMTVMO surpasses that of recently reported high-entropy DRX cathodes, underscoring its strong potential as a next-generation lithium-ion battery cathode. Collectively, these results demonstrate that the rational combination of high-entropy design and lattice-distortion-enabled percolation engineering can simultaneously achieve excellent synthetic feasibility, an exceptional Li$^+$ percolation network, and a high first-cycle capacity in DRX cathode materials.

**Table 2.** Comparison of the first-cycle capacity in different high-entropy cathodes.

| Cathodes | Capacity (mAh/g) | Ref. |
|---|---|---|
| $Li_{1.2}Mn_{0.2}Ti_{0.2}V_{0.2}Mo_{0.2}O_2$ | 256.3 | This work |
| $Li_{1.15}Mn^{3+}_{0.45}Cr_{0.1}Ti_{0.2}Mn^{4+}_{0.1}O_2$ | 233.0 | 25 |
| $Li_{1.15}Mn^{3+}_{0.45}Cr_{0.1}Ti_{0.1}Mn^{4+}_{0.2}O_2$ | 250.0 | 25 |
| $Li_{1.15}Mn^{3+}_{0.55}Al_{0.075}Ti_{0.05}Mn^{4+}_{0.1}Nb_{0.075}O_2$ | 217.0 | 25 |
| $Li_{1.3}Mn^{2+}_{0.1}Co_{0.1}Mn^{3+}_{0.1}Cr_{0.1}Ti_{0.1}Nb_{0.2}O_{1.7}F_{0.3}$ | 255.0 | 26 |

## 2.5 Short-order range analysis.

The distribution of tetrahedral clusters in all DRX materials is not fully random due to SRO induced by interactions among TM cations[27]. The SRO impedes the connectivity of 0-TM diffusion channels, thereby restricting the percolation network within the DRX cathode[28]. Therefore, suppressing SRO is essential to mitigate this limiting effect and enhance Li$^+$ transport. The increasing occurrence of 0-TM clusters in the sequence of LMZO → LMTO → LMTVMO indicates effective SRO suppression.

Concurrently, lattice distortion subtly modifies the overall SRO, as evidenced by minor increases in the 0-TM fraction in MC-MD simulations: LMZO (1.99% → 2.21%), LMTO (4.82% → 5.42%), and LMTVMO (6.65% → 7.36%) compared to distortion-free Monte Carlo results (Fig. 5a and Supplementary Note 5). This slight rise stems from reduced Coulombic repulsion due to lattice distortion. This distortion enhances tolerance to high-valence TM aggregation, as reflected in the slightly higher fractions of 3-TM clusters (e.g., LMZO: 5.87% → 6.30%) and 4-TM clusters (0.09% → 0.19%). The distribution of TMs within 3-TM and 4-TM clusters further suggests a preference for low-valence metals, due to their lower cationic repulsion (Fig. 5b). Furthermore, the progressive increase in tolerance for high-valence TM aggregation across the LMZO → LMTO → LMTVMO demonstrates that enhanced lattice distortion actively promotes such aggregation.

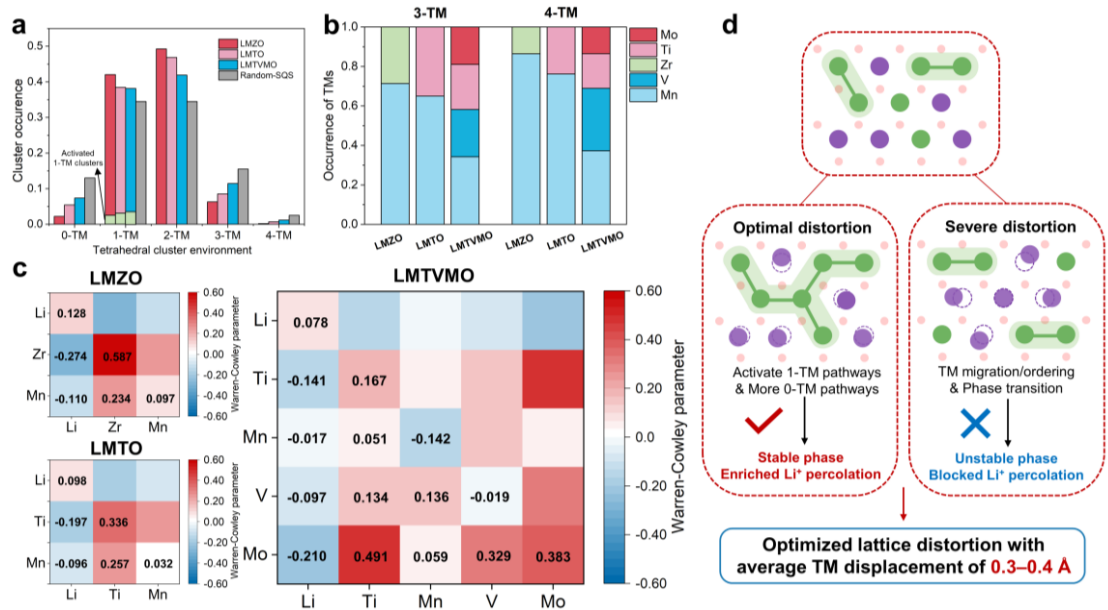

**Fig. 5** (a) Comparison of average distributions of various tetrahedral clusters (0-TM, 1-TM, 2-TM, 3-TM, and 4-TM) in MC-MD simulations converged LMZO, LMTO, and LMTVMO, with the distribution of SQS included as a fully random limit reference. (b) Distributions of 3-TM and 4-TM Clusters in three DRX Cathodes. (c) Warren–Cowley parameters of different metal species in three DRX cathodes. The color of a block visualizes the attraction (blue) or repulsion (red) between two metal species. (d) Design strategy for high-performance DRX cathodes.

A suppression of SRO is observed in LMTVMO, as evidenced by quantitative analysis of the Warren-Cowley parameters (WCPs) (Supplementary Note 6)[29]. The cation-cation interactions, both attractive and repulsive, are weakened (Fig. 5c). The Li-Li WCP decreases progressively from LMZO to LMTO to LMTVMO, indicating an increasingly random Li distribution. This progressive disordering promotes the formation of Li$_4$ tetrahedral clusters (Fig. 5a), while enhanced lattice distortion activates more 1-TM diffusion channels, thereby enhancing the Li$^+$ percolation network.

The suppression of SRO in LMTO relative to LMZO stems from the ionic radius mismatch between Zr$^{4+}$ and Ti$^{4+}$. This mismatch enhances lattice distortion, which promotes the aggregation of high-valence TMs and Li$^+$ clustering, thereby reducing the SRO. In LMTVMO, configurational entropy stabilizes the random cation distribution[24], while chemical unevenness arising from multiple TMs with differing ionic radii further intensifies lattice distortion compared with LMTO. Together, these effects lead to a pronounced suppression of SRO. To assess the influence of charge effects on SRO, we performed magnetic moments analysis based on DFT calculations (see Supplementary Note 7). We found that incorporating V and Mo increases the proportion of Mn$^{2+}$, thereby suppressing the Li$^+$/Mn$^{2+}$ clustering, as their low valence impairs local electroneutrality in the oxygen-anion sublattice. These observations are consistent with the WCP analysis (Fig. 5c), and TM valence states are further experimentally confirmed by X-ray photoelectron spectroscopy (XPS) analysis (Fig. S10). Meanwhile, the high-entropy effects introduce multiple TM valences, which promote clustering among low-valence TMs (e.g., Mn and V in Fig. 5b). Consequently, the resulting charge effect facilitates Li aggregation into Li$_4$ tetrahedra, thereby enhancing the Li$^+$ percolation in LMTVMO (Fig. 3d).

## Discussion and conclusion

In contrast to prior studies that primarily rationalize Li$^+$ transport in cation-disordered rocksalt oxides through the connectivity of idealized 0-TM diffusion channels, this work identifies lattice distortion as an essential and previously underappreciated chemical factor governing ion percolation. Rather than acting as a passive geometric perturbation, lattice distortion actively reshapes the local energy landscape, enabling Li$^+$ migration through nominally inaccessible 1-TM environments and fundamentally extending the percolation topology of disordered oxides.

Beyond directly activating additional diffusion channels, local lattice distortion exerts a profound influence on cation configurational statistics. For example, substituting Zr$^{4+}$ with Ti$^{4+}$ in Li$_{1.2}$Mn$_{0.4}$Zr$_{0.4}$O$_2$ introduces a larger ionic size mismatch, intensifying local distortions and increasing the lattice tolerance for transition-metal aggregation. This enhanced tolerance promotes the formation and connectivity of Li$_4$ tetrahedral clusters, thereby suppressing SRO and further amplifying the Li$^+$ percolation network. These findings are consistent with Ouyang's observation that off-lattice distortions are a primary driver of disorder in high-entropy disordered rocksalt cathodes[30], and we provide a mechanistic reconciliation of Ceder's reported negative correlation between average transition-metal ionic radius and Li$^+$ percolation[12], which was derived from cluster-expansion Monte Carlo models that do not explicitly account for lattice relaxation. In contrast to size-based statistical trends, distortion-enabled clustering generates energetically favorable diffusion pathways that can override purely geometric constraints.

Our results reveal that the enhancement of Li$^+$ percolation arises from a synergistic interplay between lattice distortion and configurational disorder. Using high-entropy substitution as an illustrative strategy, the incorporation of multiple transition metals (for example, V and Mo into Li$_{1.2}$Mn$_{0.4}$Ti$_{0.4}$O$_2$) plays a dual role. First, it amplifies local lattice distortions, activating a larger population of 1-TM diffusion channels while increasing the tolerance for transition-metal aggregation and Li$_4$ cluster connectivity. Second, it increases configurational entropy, favoring more random cation distributions and further suppressing SRO, as corroborated by mixing temperature calculations. Together, these effects cooperatively expand the percolating Li$^+$ network beyond what can be achieved through disorder or composition alone.

Importantly, lattice distortion is not an unbounded design parameter. Excessive distortion can destabilize the rocksalt framework, promote ordering during electrochemical cycling, trigger large-scale TM migration, and ultimately induce a phase transition that degrades electrochemical performance (Fig. 5d). Molecular dynamics simulations performed over a range of temperatures reveal a critical threshold beyond which the mean square displacement of transition-metal ions increases abruptly, signaling the onset of structural instability (Supplementary Note 8). Based on these observations, we identify an optimal distortion regime that optimizes lattice distortion with an average TM displacement of 0.3–0.4 Å, enabling enhanced Li$^+$ percolation while preserving structural integrity. This establishes a quantitative chemical guideline for balancing transport enhancement against lattice stability in disordered oxides.

In summary, this work establishes lattice distortion as an active chemical control parameter for ion percolation in cation-disordered solids. By integrating machine-learning-accelerated hybrid Monte Carlo–molecular dynamics simulations with targeted experimental validation, we introduce a lattice-responsive and quantitatively predictive framework for analyzing transport in complex disordered structures. The successful design and synthesis of the Li-rich high-entropy oxide Li$_{1.2}$Mn$_{0.2}$Ti$_{0.2}$V$_{0.2}$Mo$_{0.2}$O$_2$, exhibiting enhanced lattice distortion, expanded Li$^+$ percolation, and close agreement between predicted and measured capacity, demonstrates the practical power of this approach. More broadly, these findings revise the prevailing static percolation paradigm and suggest that ion transport in disordered solids is governed by a dynamic coupling between local chemistry, lattice response, and configurational order. The principles elucidated here are general and extend beyond lithium-ion cathodes to a wide range of ion-conducting materials, including sodium-ion conductors, oxide-ion conductors, and mixed-anion solids, offering new opportunities for the rational chemical design of functional disordered materials.

## 3. Methods

**DFT calculation.** Spin-polarized density functional theory (DFT) calculations are performed utilizing the Vienna ab initio simulation package (VASP)[31]. The Perdew-Burke-Ernzerhof (PBE) functional[32] is used for exchange-correlation energy, and van der Waals interactions are also included through the DFT-D3 method[33] with the Becke-Johnson damping function. The projector-augmented wave (PAW) method[34–36] is adopted to describe the core electron wave functions with a cutoff energy of 520 eV. To correct the DFT self-interaction error, the Hubbard-U correction is employed for the TMs $d$ states (3.9 eV for Mn, 3.25 eV for V, and 4.38 eV for Mo)[37]. The electronic energies and forces are converged to $1.0\times10^{-4}$ eV and $2.0\times10^{-2}$ eV/Å, respectively.

**Special quasi-random structure and mixing temperature calculation.** Special quasi-random structures (SQSs)[26], which describe the small-scale atomic distribution in random structures, are used to model the random cation distribution in high-entropy DRX. SQSs with the composition of $Li_{1.2}Mn_{0.2}Ti_{0.2}M^{(I)}_{0.2}M^{(II)}_{0.2}O_2$ are generated, where M1 and M2 represent different TMs. The mixing temperature[6,24] of DRX materials proposed by Ceder's group, is calculated by the following equation:

$$T_{\text{mixing}} = \frac{E_{\text{hull}}}{-k_B(0.6 ln 0.6 + \sum_M x_M ln x_M)}$$

where $E_{\text{hull}}$ is the energy above the convex hull per cation, $x_M$ refers to the atomic fraction of metal $M$ per cation site and $k_B$ is the Boltzmann constant.

**Nudged elastic band calculations.** The climbing-image nudged elastic band (CI-NEB) method[38] is used to compute Li$^+$ migration barriers, based on five linearly interpolated images along the diffusion paths. The energies and forces are converged to $1.0\times10^{-4}$ eV and $5.0\times10^{-2}$ eV/Å, respectively. The divacancy mechanism[39] is utilized to evaluate the Li$^+$ diffusion barrier in different tetrahedral clusters, where Li$^+$ diffuses between two octahedral ($o$) sites via an intermediate tetrahedral ($t$) site ($o$-$t$-$o$ diffusion)[5]. The kinetic resolved activation (KRA) barrier[40], which is independent of hop direction, is defined by the following equation:

$$\Delta E_{\text{KRA}} = E_t - \frac{1}{2}(E_i + E_f)$$

where $E_t$, $E_i$, $E_f$ is the energy of the transition state, initial state, and final state, respectively.

**Machine learning force field obtained by fine-tuning CHGNet.** Crystal Hamiltonian

Graph Neural Network (CHGNet)[21], a graph neural network based machine-learning interatomic potential with explicit inclusion of magnetic moments, is utilized to accelerate simulations. To obtain accurate force fields, three distinct MLFFs are trained via fine-tuning based on CHGNet for LMZO, LMTO, and LMTVMO, respectively. The DFT dataset used for fine-tuning is generated using an enumeration algorithm implemented in Python Materials Genomics[41], retaining the optimized trajectories from DFT relaxation. The dataset (relaxation trajectory configurations) is split into training, validation, and test sets with a ratio of 8:1:1, and the learning rate is set to be 0.001.

**Monte Carlo-Molecular Dynamics simulations and percolation analysis.** Three distinct SQSs with a $3\sqrt{5} \times 3\sqrt{5} \times 6$ supercell with 2160 atoms are constructed for LMZO, LMTO, and LMTVMO, respectively. After fine-tuning, MLFFs accelerated Monte Carlo (MLFF-MC) simulations using the Metropolis-Hastings algorithm are performed for SQSs at 1273.15 K, randomly swapping different cation atoms to obtain MC-converged supercells[12]. To introduce the lattice distortion and obtain more realistic configurations, a hybrid Monte Carlo-Molecular Dynamics (MC-MD) simulation is performed by alternatively applying 10 MC steps and 10 MD steps using Large-scale Atomic/Molecular Massively Parallel Simulator (LAMMPS) software[42], starting from a converged MC configuration. The MD step is conducted at 300 K in the canonical NVT ensemble with a time step of 1 fs. The cation tetrahedral clusters can be categorized by the number of neighboring sites occupied by Li ions, including 0-TM, 1-TM, 2-TM, 3-TM, and 4-TM[23]. The distributions of cation tetrahedral clusters are analyzed by averaging over 500 converged atomic configurations. All 0-TM $Li^+$ sites and 1-TM $Li^+$ sites with cluster tetrahedron heights above the threshold are identified as candidate percolation sites. The Dribble package[10] is then used to determine the percolating network formed by these sites and to compute the accessible capacity.

**Percolation ratio analysis and capacity evaluation.** To quantitatively evaluate the percolating $Li^+$ in three cathodes, a connectivity analysis is performed by calculating

the fraction of percolating Li relative to the total Li content, using the following formula:

$$P(s) = \frac{\sum_{i=1}^{N} I(s_i) = 1}{N} \times 100\%$$

where $N$ is the total number of Li sites, $I(s_i)$ is the indicator function to record the percolation sites ranging from $i = 1$ to $N$, and $s$ is the percolating species (e.g., Li). The capacity of the cathode based on the percolation ratio can be expressed as:

$$C = \frac{nF}{3.6m} \times P(s)$$

where $n$ represents the amount of charge transfer, $F$ is Faraday's constant, and $m$ is the relative molecular mass of the cathode material.

**Material synthesis.** All compounds were synthesized using a one-pot, room-temperature mechanochemical method. Stoichiometric amounts of $Li_2O$, $Mn_2O_3$, $TiO_2$, $V_2O_3$, and $MoO_2$ (all from Sigma-Aldrich, 99% purity) were directly mixed and processed in a planetary ball mill (FRITSCH, Pulverisette 7). Approximately 1.8 g of the powder mixture was loaded into a 45 mL zirconia milling jar together with 5 mm zirconia balls, maintaining a powder-to-ball weight ratio of 1:30. All sample handling was carried out in an argon-filled glovebox with $H_2O$ and $O_2$ levels below 0.01 ppm. High-energy ball milling was performed at 600 rpm for 30 h, employing a cyclic protocol of 30 min milling followed by 30 min rest. Notably, no post-synthesis heat treatment was required.

**Materials characterizations.** The morphology was examined using Scanning Electron Microscopy (SEM, model Hitachi SU-70), while Transmission Electron Microscopy (TEM, model JEOL JEM-F200) in conjunction with Energy Dispersive Spectroscopy (EDS) provided insights into lattice fringe and elemental distribution. Synchrotron XRD patterns of the as-synthesized compounds were obtained at Beamline BL13SSW at the Shanghai Synchrotron Radiation Facility. The XRD patterns were refined using FullProf Suite software.

**Electrochemical test.** The synthesized cathode material was mixed with Super C65

(Timcal) and polyacrylonitrile (PAN) at a mass ratio of 7:2:2 using N-methyl-2-pyrrolidone (NMP) as the solvent. The resulting slurry was uniformly cast onto aluminum foil and dried in a vacuum oven at 120 °C for 12 h. The prepared cathode electrodes exhibited an areal mass loading of approximately 1.5–2.0 mg cm$^{-2}$. CR2025 coin-type half-cells were assembled in an argon-filled glovebox with $H_2O$ and $O_2$ levels maintained below 0.01 ppm to evaluate electrochemical performance. Lithium metal served as the counter electrode, and a glass-fiber membrane was used as the separator. Commercial 1M $LiPF_6$ in an ethylene carbonate and dimethyl carbonate solution (1:1 volume ratio) was used as the electrolyte. Galvanostatic charge-discharge measurements were performed at room temperature on a LANHE battery testing system over a voltage window of 1.5-4.7 V.

## ACKNOWLEDGMENTS

This work was partially supported by grants from the National Key Research and Development Program of China (2022YFB2502100) and the National Natural Science Foundation of China (92372112, 22373005, 52573249). The calculations were supported by the High-performance Computing Platform of Peking University and Eastern Institute of Technology, Ningbo.

## References

1. Yabuuchi, N. *et al.* High-capacity electrode materials for rechargeable lithium batteries: $Li_3NbO_4$-based system with cation-disordered rocksalt structure. *Proc. Natl. Acad. Sci. U.S.A.* **112**, 7650–7655 (2015).

2. Lee, J. *et al.* Unlocking the Potential of Cation-Disordered Oxides for Rechargeable Lithium Batteries. *Science* **343**, 519–522 (2014).

3. Urban, A., Lee, J. & Ceder, G. The Configurational Space of Rocksalt-Type Oxides for High-Capacity Lithium Battery Electrodes. *Advanced Energy Materials* **4**, 1400478 (2014).

4. Goodenough, J. B. Evolution of Strategies for Modern Rechargeable Batteries. *Acc. Chem. Res.* **46**, 1053–1061 (2013).

5. Zhang, Z., Du, P.-H., Liu, J., Xia, D. & Sun, Q. Mechanisms of Enhanced Electrochemical Performance by Chemical Short-Range Disorder in Lithium Oxide Cathodes. *ACS Nano* **19**, 6554–6562 (2025).

6. Urban, A., Matts, I., Abdellahi, A. & Ceder, G. Computational Design and Preparation of Cation-Disordered Oxides for High-Energy-Density Li-Ion Batteries. *Advanced Energy Materials* **6**, 1600488 (2016).

7. Richards, W. D., Dacek, S. T., Kitchaev, D. A. & Ceder, G. Fluorination of Lithium-Excess Transition Metal Oxide Cathode Materials. *Advanced Energy Materials* **8**, 1701533 (2018).

8. Kitchaev, D. A. *et al.* Design principles for high transition metal capacity in disordered rocksalt Li-ion cathodes. *Energy Environ. Sci.* **11**, 2159–2171 (2018).

9. Seo, D.-H. *et al.* The structural and chemical origin of the oxygen redox activity in layered and cation-disordered Li-excess cathode materials. *Nature Chem* **8**, 692–697 (2016).

10. Urban, A. Modeling ionic transport and disorder in crystalline electrodes using percolation theory. Preprint at http://arxiv.org/abs/2302.06759 (2023).

11. Lee, J. *et al.* A new class of high capacity cation-disordered oxides for rechargeable lithium batteries: Li–Ni–Ti–Mo oxides. *Energy Environ. Sci.* **8**, 3255–3265 (2015).

12. Ji, H. *et al.* Hidden structural and chemical order controls lithium transport in cation-disordered oxides for rechargeable batteries. *Nat Commun* **10**, 592 (2019).

13. Wang, S. *et al.* Design principles for sodium superionic conductors. *Nat Commun* **14**, 7615 (2023).

14. Wang, S., Liu, Y. & Mo, Y. Frustration in Super-Ionic Conductors Unraveled by the Density of Atomistic States. *Angew Chem Int Ed* **62**, e202215544 (2023).

15. Sun, Y. *et al.* Expandable Li Percolation Network: The Effects of Site Distortion in Cation-Disordered Rock-Salt Cathode Material. *J. Am. Chem. Soc.* **145**, 11717–11726 (2023).

16. Zunger, A., Wei, S.-H., Ferreira, L. G. & Bernard, J. E. Special quasirandom structures. *Phys. Rev. Lett.* **65**, 353–356 (1990).

17. Yao, Y. *et al.* High-entropy Mg1.8R0.2Al4Si5O18 (R = Ni, Co, Zn, Cu, Mn) cordierite ceramics: Influence of octahedral distortion and electronegativity mismatch on the microwave dielectric properties. *Ceramics International* **50**, 51826–51831 (2024).

18. Aidhy, D. S. Chemical randomness, lattice distortion and the wide distributions in the atomic level properties in high entropy alloys. *Computational Materials Science* **237**, 112912 (2024).

19. Jun, K. *et al.* Lithium superionic conductors with corner-sharing frameworks. *Nat. Mater.* **21**, 924–931 (2022).

20. Wang, H. *et al.* Multifunctional High Entropy Alloys Enabled by Severe Lattice Distortion. *Advanced Materials* **36**, 2305453 (2024).

21. Deng, B. *et al.* CHGNet as a pretrained universal neural network potential for charge-informed atomistic modelling. *Nat Mach Intell* **5**, 1031–1041 (2023).

22. Zeng, Y. *et al.* High-entropy mechanism to boost ionic conductivity. *Science* **378**, 1320–1324 (2022).

23. Zhang, Z., Liu, J., Du, P.-H., Xia, D. & Sun, Q. Screening Na-Excess Cation-Disordered Rocksalt Cathodes with High Performance. *ACS Nano* **18**, 30584–30592 (2024).

24. Lun, Z. *et al.* Cation-disordered rocksalt-type high-entropy cathodes for Li-ion batteries. *Nat.*

*Mater.* **20**, 214–221 (2021).

25. Liu, F. *et al.* Design Principles for High-Capacity, Long-Life Cation-Disordered Rocksalt Cathodes. *Adv Funct Materials* **35**, 2419907 (2025).

26. Zhou, B. *et al.* Improved Performance of High-Entropy Disordered Rocksalt Oxyfluoride Cathode by Atomic Layer Deposition Coating for Li-Ion Batteries. *Small Structures* **5**, 2400005 (2024).

27. Squires, A. G. & Scanlon, D. O. Understanding the limits to short-range order suppression in many-component disordered rock salt lithium-ion cathode materials. *J. Mater. Chem. A* **11**, 13765–13773 (2023).

28. Clément, R. J., Lun, Z. & Ceder, G. Cation-disordered rocksalt transition metal oxides and oxyfluorides for high energy lithium-ion cathodes. *Energy Environ. Sci.* **13**, 345–373 (2020).

29. Chen, S. *et al.* Simultaneously enhancing the ultimate strength and ductility of high-entropy alloys via short-range ordering. *Nat Commun* **12**, 4953 (2021).

30. Wang, L. *et al.* Origin of enhanced disorder in high entropy rocksalt type Li-ion battery cathodes. *EES Batteries* 10.1039.D5EB00104H (2025) doi:10.1039/D5EB00104H.

31. Kresse, G. & Furthmüller, J. Efficient iterative schemes for *ab initio* total-energy calculations using a plane-wave basis set. *Phys. Rev. B* **54**, 11169–11186 (1996).

32. Kresse, G. & Joubert, D. From ultrasoft pseudopotentials to the projector augmented-wave method. *Phys. Rev. B* **59**, 1758–1775 (1999).

33. Grimme, S., Antony, J., Ehrlich, S. & Krieg, H. A consistent and accurate *ab initio* parametrization of density functional dispersion correction (DFT-D) for the 94 elements H-Pu. *The Journal of Chemical Physics* **132**, 154104 (2010).

34. Blöchl, P. E. Projector augmented-wave method. *Phys. Rev. B* **50**, 17953–17979 (1994).

35. Perdew, J. P. & Wang, Y. Accurate and simple analytic representation of the electron-gas correlation energy. *Phys. Rev. B* **45**, 13244–13249 (1992).

36. Kresse, G. & Furthmüller, J. Efficiency of ab-initio total energy calculations for metals and semiconductors using a plane-wave basis set. *Computational Materials Science* **6**, 15–50 (1996).

37. Wang, L., Maxisch, T. & Ceder, G. Oxidation energies of transition metal oxides within the GGA + U framework. *Phys. Rev. B* **73**, 195107 (2006).

38. Henkelman, G., Uberuaga, B. P. & Jónsson, H. A climbing image nudged elastic band method for finding saddle points and minimum energy paths. *The Journal of Chemical Physics* **113**, 9901–9904 (2000).

39. Ouyang, B. *et al.* Effect of Fluorination on Lithium Transport and Short-Range Order in Disordered-Rocksalt-Type Lithium-Ion Battery Cathodes. *Advanced Energy Materials* **10**, 1903240 (2020).

40. Guo, X., Chen, C. & Ong, S. P. Intercalation Chemistry of the Disordered Rocksalt $Li_3V_2O_5$ Anode from Cluster Expansions and Machine Learning Interatomic Potentials. *Chem. Mater.* **35**, 1537–1546 (2023).

41. Ong, S. P. *et al.* Python Materials Genomics (pymatgen): A robust, open-source python library for materials analysis. *Computational Materials Science* **68**, 314–319 (2013).

42. Thompson, A. P. *et al.* LAMMPS - a flexible simulation tool for particle-based materials modeling at the atomic, meso, and continuum scales. *Computer Physics Communications* **271**, 108171 (2022).

Supporting Information of

# Coupling Lattice Distortion and Cation Disorder to Control Li-ion Transport in Cation-Disordered Rocksalt Oxides

Zichang Zhang[1,#], Lihua Feng[2,#], Jiewei Cheng[1], Peng-Hu Du[1], Chu-Liang Fu[2], Jian Peng[2,*], Shuo Wang[2,*], Dingguo Xia[1,3], Xueliang Sun[2], and Qiang Sun[1,3,*]

[1]*School of Materials Science and Engineering, Peking University, Beijing, 100871, China*

[2]*Zhejiang Key Laboratory of All-Solid-State Battery, Eastern Institute of Technology, Ningbo, 315200, China.*

[3]*Beijing Key Laboratory of Theory and Technology for Advanced Batteries Materials, Peking University, Beijing 100871, China*

*Corresponding author   *Email address:* sunqiang@pku.edu.cn (Q. Sun); shuowang@eitech.edu.cn (S. Wang); jpeng@eitech.edu.cn (J. Peng)

[#]The authors (Z. C. Zhang and L. H. Feng) contribute equally to this work.

## Supplementary Note 1: Limitation of n-TM(n≥2) diffusion channels

The diffusion channels of 2-TM, 3-TM, and 4-TM are negligible owing to their substantially higher barriers. For instance, in a typical DRX cathode $Li_{1.2}Mn_{0.4}Ti_{0.4}O_2$, the barriers for 2-TM (Mn) and 2-TM (Ti) are calculated to be 1.02 eV and 1.62 eV (see Fig. S1), which significantly hinder $Li^+$ diffusion in the 2-TM environment. Besides, 3-TM and 4-TM tetrahedron clusters don't have $Li^+$ diffusion channels.

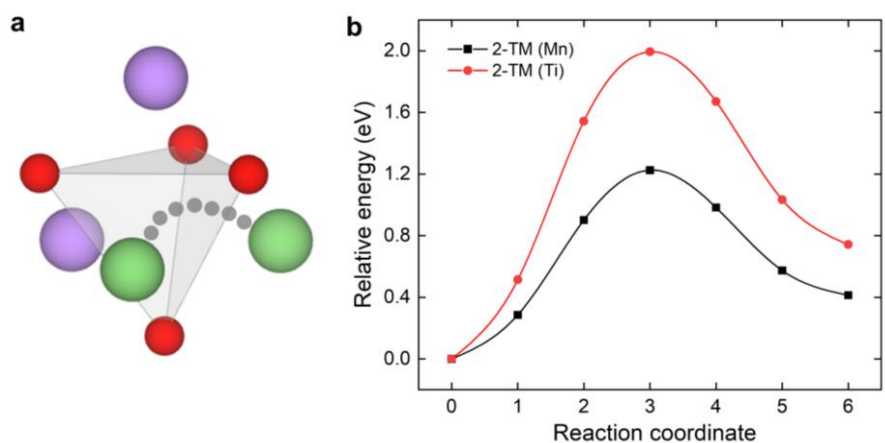

**Fig. S1. The $Li^+$ diffusion in the 2-TM tetrahedron cluster.** (a) The schematic image of $Li^+$ diffusion in the 2-TM cluster. (b) The barriers of $Li^+$ migration in the 2-TM (Mn) and 2-TM (Ti) of $Li_{1.2}Mn_{0.4}Ti_{0.4}O_2$.

## Supplementary Note 2: Models, Machine learning force field training, MC, and hybrid MC-MD simulations

To model the structures of $Li_{1.2}Mn_{0.4}Zr_{0.4}O_2$ (LMZO) and $Li_{1.2}Mn_{0.4}Ti_{0.4}O_2$ (LMTO), special quasi-random structures (SQSs)[1] with 2160 atoms are generated to represent their fully cation-disordered configuration. MLFF is utilized to accelerate the simulation to obtain the thermodynamically stable configurations of LMZO and LMTO. Accurate MLFFs are obtained by fine-tuning the universal neural network potential CHGNet using DFT-relaxed trajectory configurations[2], which is an effective paradigm for adapting general models to specific downstream tasks[3]. The DFT dataset used in CHGNet fine-tuning is shown in Table S1. As shown in Fig. S2, the fine-tuned model demonstrates excellent agreement with DFT, yielding energy mean absolute errors (MAEs) of 1.13 meV/atom for LMZO and 1.17 meV/atom for LMTO, with corresponding force MAEs of 9.61 meV/Å and 9.77 meV/Å, respectively. Meanwhile, the MLFF for LMTVMO also demonstrates excellent agreement with DFT calculations, where the MAEs for energy and force are 2.87 meV/atom and 13.5 meV/Å, respectively. These slightly higher errors compared with LMZO and LMTO are attributed to the increased complexity of the high-entropy DRX cathode structure. Using high-accuracy MLFFs, MC simulations are conducted on SQSs at 1273.15 K through randomly swapping one pair of cation atoms in each step, yielding MC-converged configurations that reach thermodynamic equilibrium. As shown in Fig. S3, the energies of LMZO and LMTO converge after 100000 MC steps, while that of LMTVMO converges after 200000 MC steps, reflecting the increased configurational complexity of the high-entropy DRX cathode. Since MLFF-MC and CE-MC simulations are typically performed on ideal lattices without structural relaxation, the local lattice distortions and displacive effects are inherently ignored, which has been demonstrated to be significant in capturing features of SRO[4]. To introduce the local lattice distortions, a hybrid Monte Carlo-Molecular Dynamics (MC-MD) simulation is conducted on the above cathode materials. This approach simultaneously samples configurational and displacive degrees of freedom, thereby achieving more realistic thermodynamic equilibrium configurations after energies converge. Starting from the MC-converged configurations, hybrid MC-MD simulations are conducted by alternating 10 fs of MD simulations with 10 MC steps. As shown in Fig. S4, LMZO, LMTO, and LMTVMO achieve equilibrium after a sufficient period of simulation.

**Table S1.** Scale of the DFT dataset used for CHGNet fine-tuning. (Unit: number of configurations)

| Dataset | LMZO | LMTO | LMTVMO |
| --- | --- | --- | --- |
| Initial structures | 476 | 476 | 512 |
| Relaxation trajectory configurations | 7572 | 8991 | 13823 |

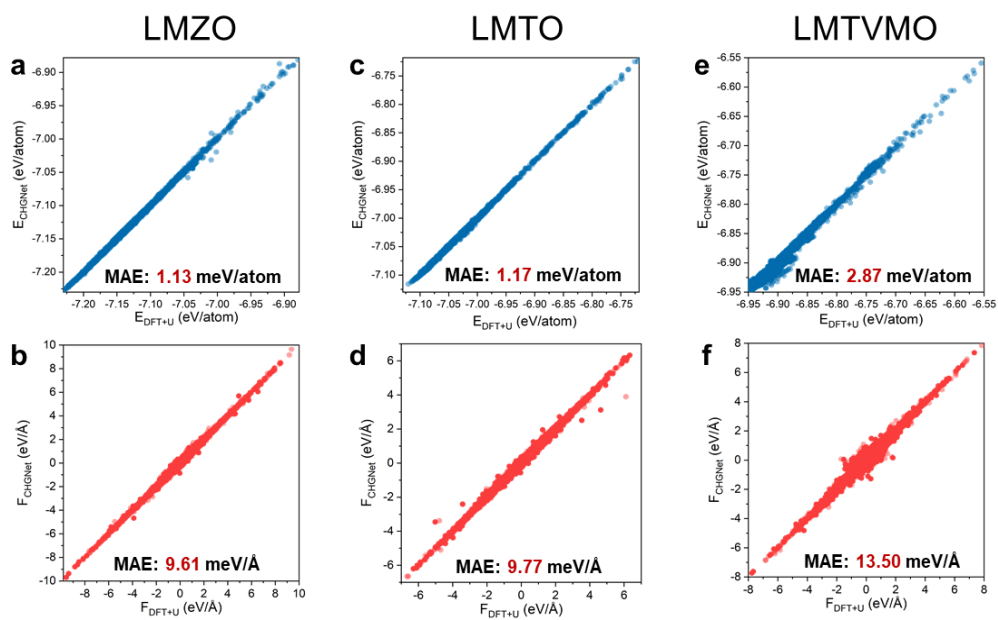

**Fig. S2.** Comparison between DFT+U calculations and fine-tuned CHGNet predictions for (a) energies and (b) forces in LMZO, (c) energies and (d) forces in LMTO, and (e) energies and (f) forces in LMTVMO.

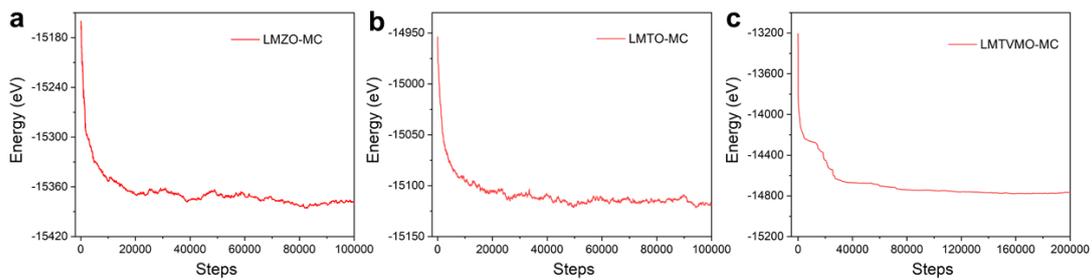

**Fig. S3.** Monte Carlo simulation in Metropolis−Hastings algorithm with 100000 steps in (a) LMZO and (b) LMTO, and 200000 steps in (c) LMTVMO.

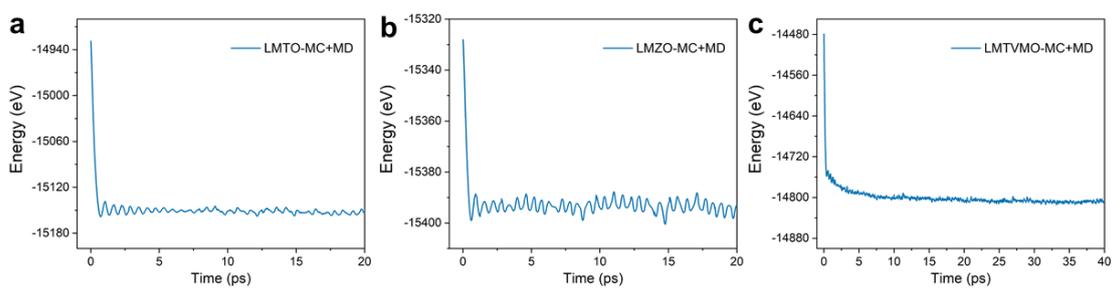

**Fig. S4.** Hybrid Monte Carlo-Molecular Dynamics simulations with 20 ps in (a) LMZO and (b) LMTO, and 40 ps in (c) LMTVMO.

# Supplementary Note 3: Visualization of the percolation network of three cathodes after MC simulation

The Li$^+$ percolation networks obtained from conventional MC simulations are illustrated in Fig. S5, where LMZO exhibits no percolation, and LMTO displays a higher percolation ratio. This is consistent with the cluster-expansion Monte Carlo (CE-MC) simulations based on the 0-TM rule in the previous report[5], and the discrepancy of the percolation networks has been explained by short-range order (SRO). Furthermore, the percolation network of LMTVMO is further enhanced compared with LMTO, indicating the further suppression of SRO after the formation of the high-entropy DRX cathode.

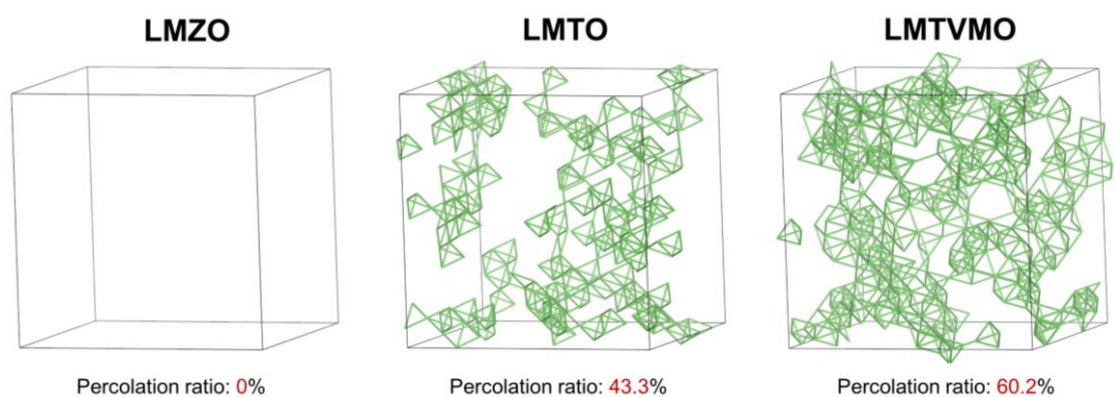

**Fig. S5.** Percolation network of representative LMZO, LMTO, and LMTVMO based on 0-TM percolation rule without distortion.

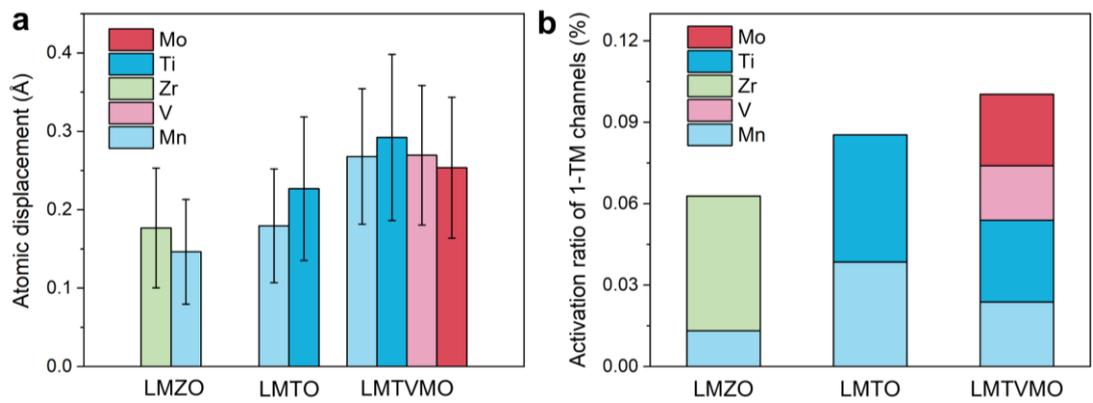

**Fig. S6.** (a) Average atomic displacement of TMs in LMZO, LMTO, and LMTVMO after hybrid MC-MD simulations, with standard deviation represented by an error bar. (b) The activation ratio of 1-TM channels upon introduction of local lattice distortion after hybrid MC-MD simulations, contribution of 1-TM channels from different TMs is labeled in different colors.

# Supplementary Note 4: Diffusion barriers of 1-TM diffusion channels in LMTVMO

To evaluate the percolation network of LMTVMO, the diffusion properties of 1-TM diffusion channels need to be determined. As the valence of $Ti^{4+}$ doesn't change, while the valence of $Mn^{3+}$ doesn't change so much after V and Mo doping, the tetrahedral height threshold in 1-TM (Mn) and 1-TM (Ti) is approximately regarded as the same in LMTVMO.

The average diffusion barriers for 1-TM (V) and 1-TM (Mo) channels are determined by sampling six different diffusion channels to obtain the average barriers in each environment. As shown in Fig. S7(a), the 1-TM (V) and 1-TM (Mo) are calculated to be 0.42 eV and 0.54 eV, respectively. To determine the tetrahedral heights threshold for low-barrier 1-TM (V) and 1-TM (Mo) diffusion channels, the variation of diffusion barriers with tetrahedral cluster height is presented in Fig. S7(b), where the critical heights for these channels are identified as 2.479 Å for V and 2.534 Å for Mo.

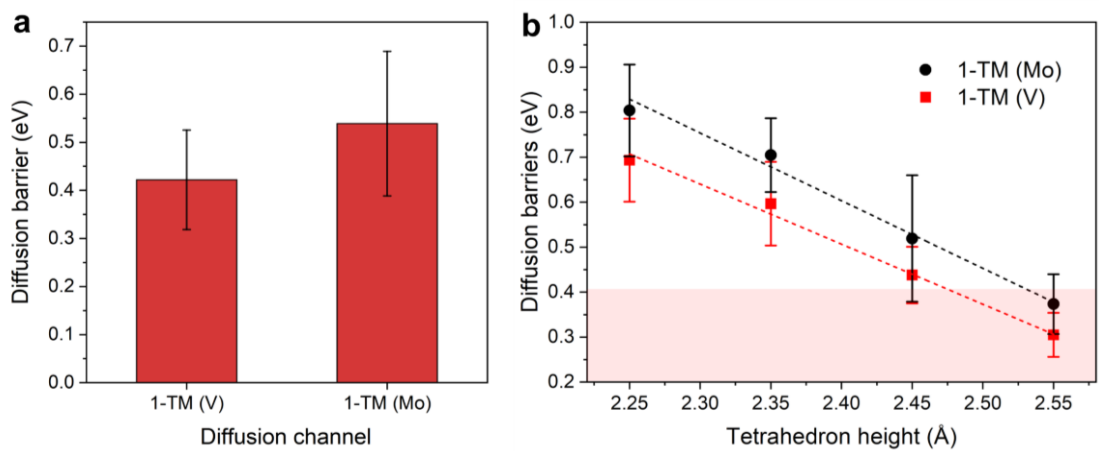

**Fig. S7.** (a) Distribution of average $Li^+$ diffusion barriers in different 1-TM environments of V and Mo in LMTVMO, standard deviation represented by an error bar. (b) The relationship between diffusion barriers and tetrahedron height in 1-TM environments of V and Mo, with the standard deviation represented by an error bar. The scatterings of Ti are moved slightly along the fitted line for clarity, and the area of diffusion barriers below 0.4 eV is labeled in pink.

Table S2. Detailed structural information of LMTVMO from XRD Rietveld refinement.

| Space group: Fm-3m | | $a=b=c=4.16040$ Å | | | volume=72.012 Å$^3$ | |
|---|---|---|---|---|---|---|
| atom | site | x | y | z | occ. | Fraction |
| Li | 4a | 0 | 0 | 0.5 | 0.01248 | 0.59914 |
| Mn | 4a | 0 | 0 | 0 | 0.00208 | 0.09986 |
| Ti | 4a | 0 | 0 | 0 | 0.00208 | 0.09986 |
| V | 4a | 0 | 0 | 0 | 0.00208 | 0.09986 |
| Mo | 4a | 0 | 0 | 0 | 0.00208 | 0.09986 |
| O | 4b | 0.5 | 0.5 | 0.5 | 0.02083 | 1 |

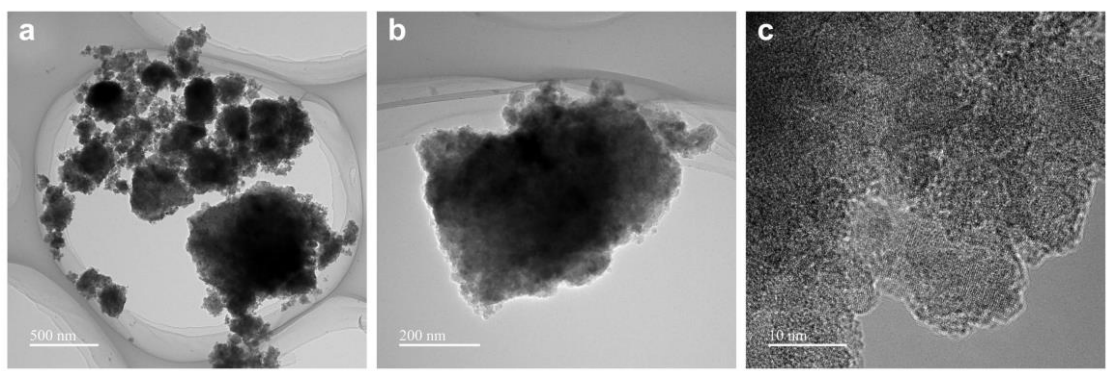

**Fig. S8.** HRTEM images of LMTVMO.

# Supplementary Note 5: Average distributions of various tetrahedral clusters in MC-converged configurations in three cathodes

The average distributions of tetrahedral clusters in MC-converged LMZO, LMTO, and LMTVMO are illustrated in Fig. S8. The ratio of 0-TM cluster increases, indicating progressively reduced SRO and improved 0-TM $Li^+$ percolation.

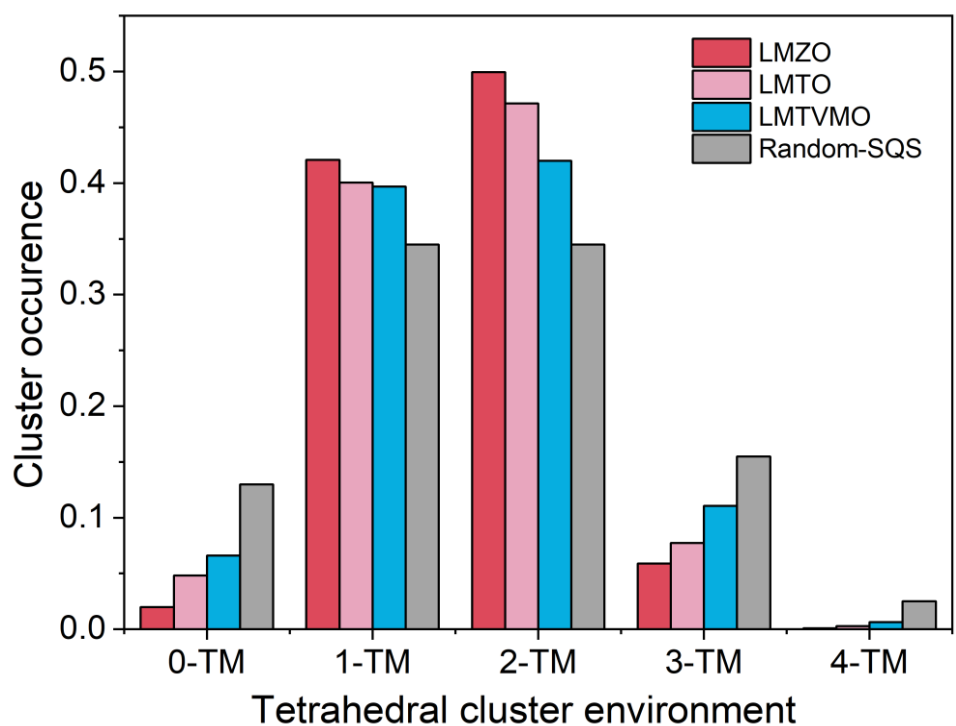

**Fig. S9.** Comparison of average distributions of various tetrahedral clusters (0-TM, 1-TM, 2-TM, 3-TM, and 4-TM) in MC simulations converged LMZO, LMTO, and LMTVMO, with the distribution of SQS included as a fully random limit reference.

## Supplementary Note 6: The Warren-Cowley parameter analysis

To elucidate the underlying mechanism of increased 0-TM clusters and reduced 1-TM clusters, the Warren-Cowley parameter (WCP)[6] is employed to characterize the first-nearest neighbor cation shell in three cathodes using the following formula:

$$\alpha_{ij} = 1 - \frac{P_{ij}(r)}{c_j}$$

where $P_{ij}(r)$ is the average probability of finding a $j$-type atom around an $i$-type atom, and $c_j$ is the overall concentration of $j$-type atoms in the system. A negative $\alpha_{ij}$ indicates a tendency for $j$-type to cluster around $i$-type atoms, while a positive value denotes the opposite. Specifically, $\alpha_{ij}$ = 0 corresponds to a random distribution. The WCP values for different metal species in LMZO, LMTO, and LMTVMO, obtained from hybrid MC-MD converged configurations, are shown in Fig. 5. It's obvious that substituting Zr with Ti weakens Li–Mn and Li–Zr interactions despite their identical valence, while the subsequent introduction of V and Mo further diminishes Li–Mn and Li–Ti interactions. Meanwhile, the interaction of TM-TM is also reduced in the order of LMZO, LMTO, and LMTVMO. These observations suggest that SRO is gradually reduced in the order of LMZO, LMTO, and LMTVMO, as the tendencies of both attraction and repulsion are progressively weakened.

## Supplementary Note 7: The magnetic moments and valence analysis

Previous work has attributed the reduced SRO in LMTO relative to LMZO to combined charge and size effects[5]. To further understand the more pronounced SRO reduction in high-entropy LMTVMO, the magnetic moments (μB) and valences of TMs are evaluated as shown in Tables S3-S5. Upon introducing V and Mo, the average magnetic moment of Mn increases from ~3.8 to ~4.2 μB, indicating partial reduction of $Mn^{3+}$ to $Mn^{2+}$. Meanwhile, the average valence of V is +3 to +4, whereas that of Mo is +4 to +5, which is consistent with a previous report[7] that the coexistence of $Mn^{3+}$ and $V^{3+}$ (or $Mo^{4+}$) induces charge transfer, oxidizing $V^{3+}$ (or $Mo^{4+}$) while reducing $Mn^{3+}$. These transition-metal valence states are further experimentally confirmed by X-ray photoelectron spectroscopy (XPS) analysis, as shown in Fig. S10. The increasing ratio of $Mn^{2+}$ reduces the likelihood of Li-Mn mixing, as $Li^+/Mn^{2+}$ mixing cannot maintain local electroneutrality in the DRX framework. Consequently, the charge effect upon introduction of V and Mo facilitates Li segregation into $Li_4$ tetrahedra, endowing LMTVMO with superior $Li^+$ percolation properties.

**Table S3.** Average magnetic moment ($\mu_B$) and valence of transition metals in LMZO.

| TMs | Mn | Zr |
|---|---|---|
| Average magnetic moment | 3.832 | 0.028 |
| Valence | ~3+ | ~4+ |

**Table S4.** Average magnetic moment ($\mu_B$) and valence of transition metals in LMTO.

| TMs | Mn | Ti |
|---|---|---|
| Average magnetic moment | 3.815 | 0.010 |
| Valence | ~3+ | ~4+ |

**Table S5.** Average magnetic moment ($\mu_B$) and valence states of transition metals in LMTVMO.

| TMs | Mn | Ti | V | Mo |
|---|---|---|---|---|
| Average magnetic moment | 4.203 | 0.055 | 1.741 | 1.484 |
| Valence states | 2+ ~ 3+ | ~4+ | 3+ ~ 4+ | 4+ ~ 5+ |

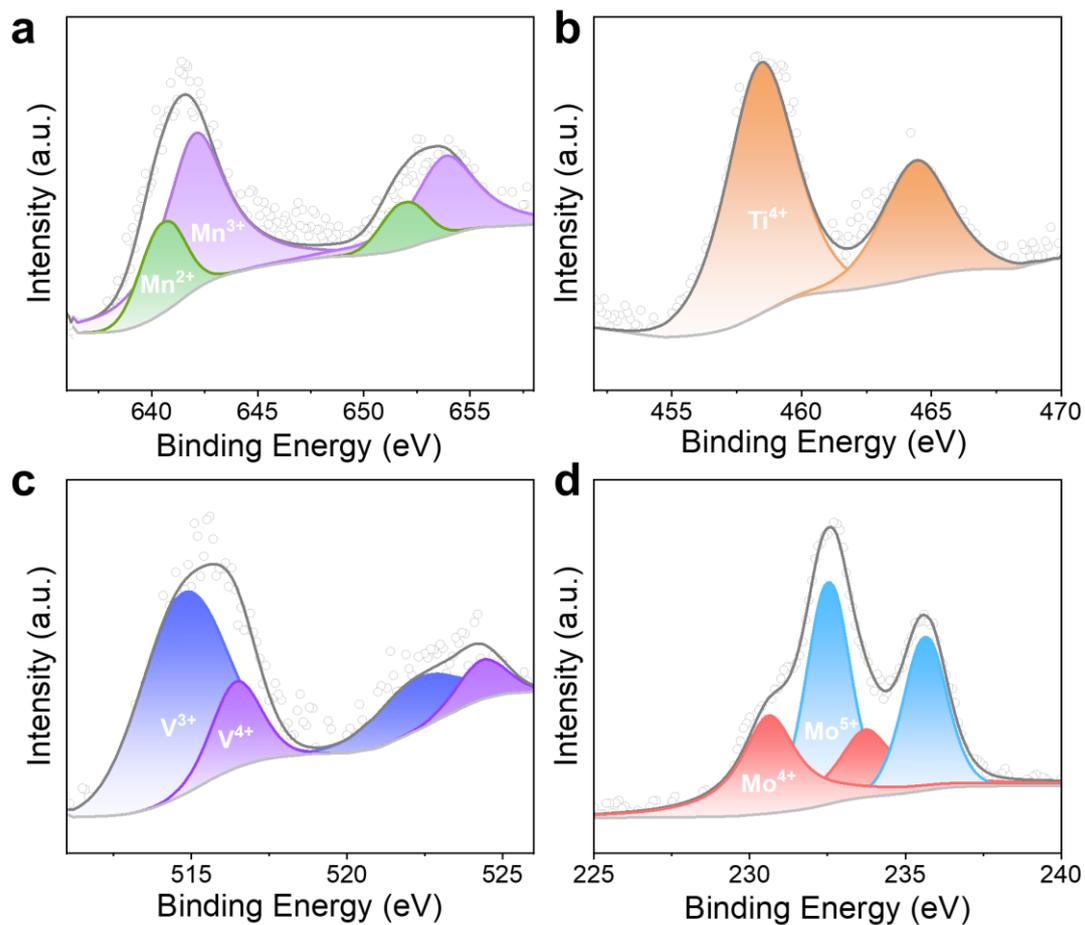

**Fig. S10.** High-resolution (a) Mn 2p, (b) Ti 2p, (c)V 2p, and (d) Mo 3d XPS spectra of LMTVMO.

# Supplementary Note 8: Critical TM displacement threshold for structural stability in DRX cathodes

To correlate the lattice distortion with structural stability, MD simulations in the NVT ensemble are performed for the three cathodes at various temperatures to obtain the mean square displacement (MSD) and the average TM displacement relative to their ideal lattice sites. As shown in Fig. S11, both the MSD and average TM displacement increase with temperature. Notably, the MSD curves exhibit clear inflection points, indicating significant structural instability, such as incipient phase transitions or large-scale TM migration, once the TM displacement exceeds these inflection points. Specifically, the inflection point corresponds to an average TM displacement of 0.35 Å for LMTO and LMZO, while that for LMTVMO occurs at 0.40 Å, higher than those of LMZO and LMTO, reflecting its enhanced tolerance to lattice distortion due to high configurational entropy[8]. A conservative stability threshold is therefore proposed, wherein the average TM displacement in DRX cathodes is maintained within 0.3–0.4 Å to enhance the Li$^+$ percolation network while preserving structural integrity.

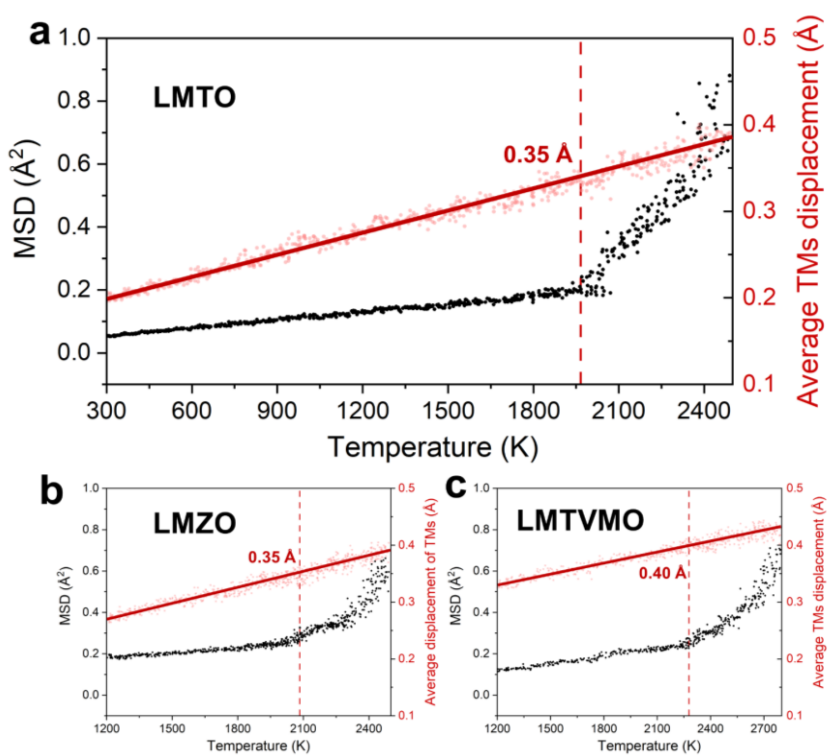

**Fig. S11.** MSD and average TM displacement as a function of temperature for the three cathodes obtained from MD simulations.

# Reference

1. Zunger, A., Wei, S.-H., Ferreira, L. G. & Bernard, J. E. Special quasirandom structures. *Phys. Rev. Lett.* **65**, 353–356 (1990).

2. Deng, B. *et al.* CHGNet as a pretrained universal neural network potential for charge-informed atomistic modelling. *Nat. Mach. Intell.* **5**, 1031–1041 (2023).

3. Chen, C. & Ong, S. P. A universal graph deep learning interatomic potential for the periodic table. *Nat. Comput. Sci.* **2**, 718–728 (2022).

4. Szymanski, N. J. *et al.* Modeling Short-Range Order in Disordered Rocksalt Cathodes by Pair Distribution Function Analysis. *Chem. Mater.* **35**, 4922–4934 (2023).

5. Ji, H. *et al.* Hidden structural and chemical order controls lithium transport in cation-disordered oxides for rechargeable batteries. *Nat. Commun.* **10**, 592 (2019).

6. Chen, S. *et al.* Simultaneously enhancing the ultimate strength and ductility of high-entropy alloys via short-range ordering. *Nat. Commun.* **12**, 4953 (2021).

7. Lun, Z. *et al.* Cation-disordered rocksalt-type high-entropy cathodes for Li-ion batteries. *Nat. Mater.* **20**, 214–221 (2021).

8. Ouyang, B. & Zeng, Y. The rise of high-entropy battery materials. *Nat. Commun.* **15**, 973 (2024).